\documentclass[]{pasj01}

\begin{document} 
\Received{}
\Accepted{}

\title{Performance test of RM CLEAN and its evaluation with chi-square value}

\author{Yoshimitsu \textsc{Miyashita}\altaffilmark{1}}
\altaffiltext{1}{Kumamoto University, 2-39-1, Kurokami, Kumamoto 860-8555, Japan}
\email{152d8016@st.kumamoto-u.ac.jp}

\author{Shinsuke \textsc{Ideguchi},\altaffilmark{2}}
\altaffiltext{2}{Department of Physics, UNIST, Ulsan 44919, Korea}

\author{Keitaro \textsc{Takahashi}\altaffilmark{1}}


\maketitle

\begin{abstract}
RM CLEAN is a standard method to reconstruct the distribution of cosmic magnetic fields and polarized sources along the line of sight (LOS) from observed polarization spectrum. This method is similar to the CLEAN algorithm for aperture synthesis radio telescope images but it is rather unclear in what cases RM CLEAN works well. In this paper, we evaluate the performance of RM CLEAN by simulating spectro-polarimetric observations of two compact sources located in the same LOS, varying the relative initial polarization angle and Faraday depth systematically. Especially, we focus on if the two polarized sources can be resolved in the Faraday depth space and how well the source parameters can be estimated. We confirm the previous studies that two sources cannot be resolved when they are closely located in the Faraday depth space for specific values of the relative initial polarization angle. Further, we calculate the chi-square value for the fit between the mock data of polarization spectrum and the one from RM CLEAN. Then we find that the chi-squared value is not always significantly large even when RM CLEAN gives wrong results.
\end{abstract}

\section{Introduction}

Cosmic magnetic fields play an important role in various astrophysical system. At small scales, they affect star formation and gas dynamics in galaxies \citep{won02,bec04,bec09b}, galactic outflows \citep{mac13} and evolution of supernova remnants (SNRs) \citep{ino13}.  At larger scales, magnetic fields are a key ingredient to understand the structure formation and evolution of galaxies \citep{hea15}, the heat conduction in the intracluster medium (ICM) as well as the radio emission from the ICM such as radio halos, radio relics and radio mini-halos in galaxy clusters \citep{fer12}. Further, magnetic fields can be a unique probe of large-scale structure of the universe by using the interaction of the high energy $\gamma$-ray in the intergalactic voids \citep{tak12,tak13} and by observing the turbulence of the cosmic web \citep{ryu08,aka10,aka11}.

One of conventional methods to probe cosmic magnetic fields is the Faraday Rotation effect, which is the rotation of polarization angle when electromagnetic waves travel through a magnetized plasma. The rotation angle is expressed as
\begin{equation}
\chi = \chi_0 + {\rm RM}~\lambda^2,
\end{equation}
where $\chi_0$ is the initial polarization angle, $\lambda$ is the wavelength and RM stands for Rotation Measure which can be written as 
\begin{equation}
{\rm RM} = 0.81 \int_{C} n_e B_{||} dx,
\end{equation}
where $n_e$ is the number density of electron in cm$^{-3}$, $B_{||}$ is the line of sight (LOS) component of the magnetic field strength in $\mu$G and $x$ is the physical distance to the source in pc. Because the rotation angle is proportional to the squared wavelength,  we can evaluate the RM value if we observe the polarization angles at multiple wavelengths. With reasonable models of electron number density such as one from X-ray observations, we can estimate the average magnetic fields strength parallel to the LOS. The method has been used for SNRs \citep{gae98}, external galaxies \citep{gae05,bec09a} and galaxy clusters \citep{fer12}.

Although the study of magnetic fields using RM has been done frequently in literature, there are two limitations in this method. One is that the linear relation between the polarization angle and squared wavelength is seen only in a simple situation with a single polarization source along the LOS. If there are multiple sources, the relation generally becomes non-linear \citep{osu12}. The other is that we can not obtain the distribution of magnetic fields and polarized sources because RM gives only an integral quantity.

A more sophisticated method to overcome these problems is the RM synthesis technique which is first proposed by \citet{bur66} and established by \citet{bre05}. This method utilize the fact that observed complex polarized intensity P($\lambda^2$) can be expressed as 
\begin{equation}
P(\lambda^2) = \int_{-\infty}^{\infty} F(\phi) e^{2i \phi \lambda^2 } d\phi.
\label{eq:P}
\end{equation}
Here $F(\phi)$ is called Faraday dispersion function (FDF) or Faraday spectrum, which represents complex polarized intensity as a function of Faraday depth $\phi$,
\begin{equation}
\phi(x) = 0.81 \int_0^x n_e B_{||} dx.
\end{equation}
Because Eq. (\ref{eq:P}) has the same form as the Fourier transform, the FDF is formally obtained by
\begin{equation}
F(\phi) = \frac{1}{2\pi}\int_{-\infty}^{\infty} P(\lambda^2) e^{-2i \phi \lambda^2 }d\lambda^2.
\end{equation}
This inverse transformation is called RM synthesis. Although Faraday depth does not generally have one-to-one correspondence with the physical distance, the FDF includes much richer information on the distribution of magnetic fields, polarization intensity and thermal electrons along the LOS, compared with the conventional RM. Thus, the technique is expected to be useful for probing a LOS structure of galaxies \citep{ide14b}, and even for probing the intergalactic fields in filaments of galaxies \citep{aka14}. Other useful methods to reconstruct the FDF include QU-fitting which is a method of model fitting without the inverse Fourier transform \citep{osu12,ide14a}, wavelet-based fitting \citep{fri11}, compressed sensing \citep{li11a,li11b} and RM MUSIC based on eigen-decomposition of the covariance matrix of the observed polarizations \citep{and13}.

\begin{figure}[t]
\begin{center}
\includegraphics[width=8cm]{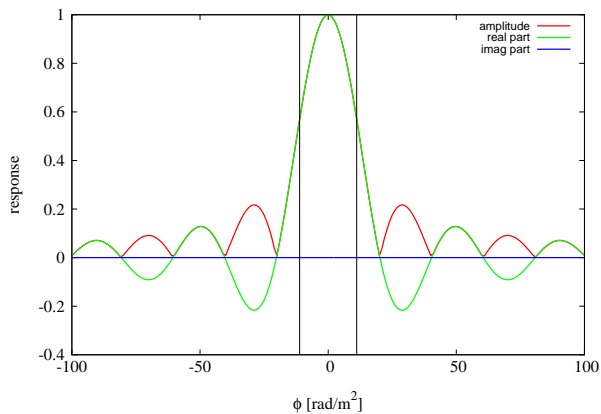} 
\end{center}
\caption{The RMSF for the ASKAP (amplitude: red line; real part: green line; imaginary part; blue line). Two black lines show the width of the FWHM for the ASKAP [22.26 rad m$^{-2}$].}
\label{fig:spread}
\end{figure}

In reality, the obtained FDF by RM synthesis is incomplete since the negative value of squared wavelength is not physical and, even for positive values of squared wavelengths, the observational data is limited by the specification of telescopes. Denoting the window function as $W(\lambda^2)$, where $W(\lambda^2) = 1$ if $\lambda^2$ is in the observable bands and otherwise $W(\lambda^2) = 0$, the inversion can be written as, 
\begin{equation}
\tilde{F}(\phi) = \frac{1}{2\pi}\int_{-\infty}^{\infty} W(\lambda^2) P(\lambda^2) e^{-2i \phi \lambda^2 } d\lambda^2,
\label{eq:F_tilde}
\end{equation}
where $\tilde{F}(\phi)$ is called the dirty FDF. Eq. (\ref{eq:F_tilde}) is rewritten using convolution as,
\begin{eqnarray}
\tilde{F}(\phi) &=& \frac{1}{K} F(\phi)*R(\phi), \\
R(\phi) &=& K \int_{-\infty}^{\infty} W(\lambda^2) e^{-2i \phi \lambda^2 } d\lambda^2, \\
K^{-1} &=& \int_{-\infty}^{\infty} W(\lambda^2) d\lambda^2.
\end{eqnarray}
Here, $R(\phi)$ is called the Rotation Measure Spread Function (RMSF), which determines the accuracy of reconstruct of the FDF, and $K$ is a normalization constant. Even if the intrinsic FDF is thin in $\phi$ space, the dirty FDF has a finite width and sidelobes due to the incomplete inverse transform. The width of the dirty FDF is estimated by the Full Width at Half Maximum (FWHM) of the RMSF,
\begin{equation}
{\rm FWHM} = \frac{2\sqrt{3}}{\Delta \lambda^2} ,~~~\Delta \lambda^2 = \lambda_{max}^2 - \lambda_{min}^2.
\end{equation}
Thus, a broadband observation is required in order to reconstruct the  FDF accurately. The Square Kilometre Array (SKA), a future project of cm-m interferometer, and its ongoing pathfinders such as Australian SKA Pathfinder (ASKAP), Murchison Widefield Array (MWA) and LOw Frequency ARray (LOFAR) can realize broadband and high sensitivity observation \citep{hea15,hav15}. For example, the value of FWHM is 22.26 [rad m$^{-2}$] and 0.189 [rad m$^{-2}$] for the ASKAP and SKA, respectively. Fig. \ref{fig:spread}  shows the amplitude, real part and imaginary part of RMSF for the ASKAP (700-1800 MHz) using the equation,
\begin{equation}
R(\phi) = K \int_{-\infty}^{\infty} W(\lambda^2) e^{-2i \phi (\lambda^2 - \lambda_0^2) } d\lambda^2. \\
\end{equation}
Here, the weighted average of the observation wavelength,
\begin{equation}
\lambda_0^2 = \frac{\int_{-\infty}^{\infty} W(\lambda^2) \lambda^2 d\lambda^2}{\int_{-\infty}^{\infty} W(\lambda^2) d\lambda^2} , \\
\label{eq:lambda0}
\end{equation}
is added to the exponential to avoid complex RMSF (see \cite{bur66} for detail). Two black lines in Fig. 1 show the width of the FWHM for the ASKAP [22.26 rad m$^{-2}$].

In order to remove the false dispersion and sidelobes of the dirty FDF, \citet{hea09} proposed RM CLEAN which is similar to the CLEAN algorithm for aperture synthesis radio telescope images \citep{hog74}. Although this works well for multiple sources sufficiently separated in $\phi$ space, \citet{far11} reported a phenomenon called RM ambiguity where false signals appear when two sources are located very closely with each other in $\phi$ space. This phenomenon is induced by the interference of two sources in $\phi$ space and, due to the RM ambiguity, two separate polarization sources cannot often be resolved, which makes the physical interpretation very difficult. Following \citet{far11}, \citet{kum14} systematically investigated the condition of the appearance of false signals, and found that the false signals depend not only on the separation but on the difference in polarization angles and the intensity ratio between two sources.

These works concern the reliability of RM CLEAN and more studies on its performance are required. In this paper, we study the effectiveness of RM CLEAN systematically, focusing on the parameter estimation of polarization sources as well as merge of two separate sources due to RM ambiguity. Following \citet{kum14}, we simulate polarization observation of two compact sources within the same sight line, varying the differences in polarization angles and Faraday depths between two sources.

Further, we investigate if the chi-square value for the fit between the observed polarization spectrum and the one obtained from RM CLEAN could be a criterion for the performance of RM CLEAN for a specific observation. In fact, \citet{sun15} performed a data challenge to evaluate how well various methods can reconstruct the FDF, and used the chi-square value as one of figures of merits for the evaluation. However, it has not been clear whether the chi-square value can be a criterion which guarantees the goodness of the reconstruction.

In section 2, we introduce RM CLEAN and describe our model and simulation method. We show the results on RM ambiguity and parameter estimation and discuss the chi-square value of the fit in section 3. Finally, we summarize the work in section 4.

\section{Model and Calculation}

\subsection{RM CLEAN}

RM CLEAN \citep{hea09} is an algorithm similar to the CLEAN deconvolution of images for radio interferometer \citep{hog74}. It removes the sidelobes of dirty FDF in order to make the physical peaks clearer and easy to identify. Here we summarize RM CLEAN briefly.

First, we seek a peak value in the $|\tilde{F}(\phi)|$ and store the peak location $\phi_p$ and the peak value $\tilde{F}(\phi_p)$ as a Faraday component. Then, we shift $R(\phi_p)$ to the location of $\tilde{F}(\phi_p)$, also set the amplitude in the same way. Secondly, we subtract the shifted-scaled RMSF $\gamma \tilde{F}(\phi) R(\phi - \phi_p)$ from $\tilde{F}(\phi)$, where $\gamma$ is a constant. Thirdly, we add a gaussian function with an amplitude of $\gamma \tilde{F}(\phi_p)$ and a width of the FWHM of the RMSF to the CLEANed FDF. Then, we repeat the above steps until $\tilde{F}(\phi_p)$ becomes below a threshold or until the iteration reaches a certain number of $N_{\rm max}$. Finally, we add the residual of the dirty FDF to the CLEANed FDF. The CLEANed FDF constructed this way is expected to be a better reconstruction of the true FDF. Finally, we define CLEAN component as a sum of Faraday components:
\begin{equation}
S(\phi) = \sum^{N_{\rm it}}_{k=1} \gamma K S^k_{\rm F}(\phi),
\end{equation}
where $S^k_{\rm F}(\phi)$ is the $k$-th Faraday component and $N_{\rm it}$ is the number of iteration. Practically, in our calculation, we set $\gamma = 0.1$ following \citet{far11}, $N_{\rm max} = 3000$, and the threshold to $0.006$ which is the noise level of dirty FDF in our simulations.

\subsection{Model}

We consider a model of FDF which consists of two delta-function sources and is written as,
\begin{equation}
F(\phi)
= f_{1} e^{2i\chi_{0,1}} \delta(\phi - \phi_1)
  + f_{2} e^{2i\chi_{0,2}} \delta(\phi -\phi_2) ,
\end{equation}
where $\phi_i$, $f_i$ and $\chi_{0,i}$ are the Faraday depth, emissivity and intrinsic polarization angle of the $i$-th source, respectively. Further, we define the difference of intrinsic polarization angles and Faraday depths as,
\begin{eqnarray}
&& \Delta \chi_0 = \chi_{0,1} - \chi_{0,2} , \\
&& \Delta \phi = \phi_2 - \phi_1 .
\end{eqnarray}

Hereafter, we fix $\phi_1 = 10~[{\rm rad/m}^2]$, $f_1 = f_2 = 10$ and $\chi_{0,2} = 0~[{\rm rad}]$, and vary $\Delta \chi_0$ ($\chi_{0,1}$) and $\Delta \phi$ ($\phi_2$) systematically to produce mock data of polarization spectrum from Eq. (\ref{eq:P}) and reconstruct FDF with RM CLEAN. From these simulations, we examine the parameter region where RM CLEAN works effectively. We consider the observation band of ASKAP (700 MHz to 1800 MHz) and set the channel width to $1~{\rm MHz}$. In producing mock data, a gaussian noise with the average 0 and variance 1 is added to each channel. This noise level results in signal-to-noise ratio of about 10 at each channel. We assume such a relatively high signal-to-noise because we would like to focus on the intrinsic performance of RM CLEAN aside from statistical fluctuations by noises. In fact, we will see qualitative features of the results do not change for larger values of $f_1$ and $f_2$ (and then signal-to-noise ratio).

\section{Results and Discussion}

\subsection{number of identified sources}

The number of polarized sources along the LOS is the most basic information to study the target. It is known that when two sources are closely located in $\phi$ space (not necessarily in physical space), false signals can appear between the two sources \citep{far11,kum14}. This phenomenon is called RM ambiguity. It happens below the resolution in $\phi$ space ($\sim$ FWHM of the RMSF) and depends on the difference of the two sources in the initial polarization angle as well. When the false signals dominate the true signals, only one source is identified in the CLEANed FDF, which makes it very difficult to understand the physical state of the source.

\begin{figure}
\begin{center}
\includegraphics[width=8cm]{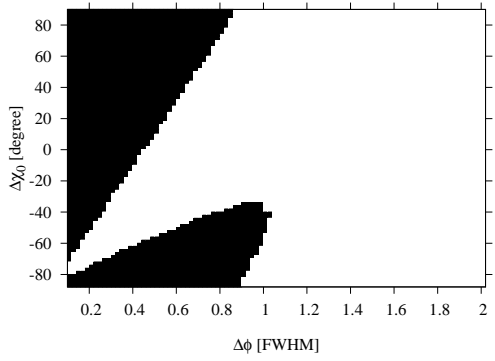} 
\end{center}
\caption{Number of sources identified with RM CLEAN in $\Delta \phi$-$\Delta \chi_0$ plane. The value of $\Delta \phi$ is in units of FWHM of the ASKAP ($\sim 22.3~{\rm rad/m^2}$). Two sources are correctly identified in the white area, while only one source is identified in the black area (RM ambiguity).}
\label{fig:sourcenum}
\end{figure}

Here, we focus on the number of identified sources in the CLEANed FDF, rather than false signals. Fig. \ref{fig:sourcenum} shows the number of identified sources in the CLEANed FDF in $\Delta \phi$-$\Delta \chi_0$ plane. Two sources are identified in the white area, while two sources are merged and only one source is identified in the black area. RM ambiguity is seen for $\Delta \phi \lesssim 1~{\rm FWHM}$ but two sources are correctly identified for as close as $\Delta \phi \sim 0.5~{\rm FWHM}$, depending the value of $\Delta \chi_0$. We can also confirm that the Fig. \ref{fig:sourcenum} has a periodicity with respect to $\Delta \chi_0$ with the period of $\pi$. This is because the polarization angle can take a value from $-\pi/2$ to $\pi/2$, and the RMSF changes its shape periodically within the range. These behaviors are consistent with the previous works \citep{far11,kum14}. Finally, we confirmed that the shape of the black area does not change for larger values (20 and 30) of $f_1$ and $f_2$.

\begin{figure*}[htb]
 \begin{center}
  \begin{tabular}{c}
   \begin{minipage}{0.5\hsize}
    \begin{center}
     \includegraphics[width=8.5cm]{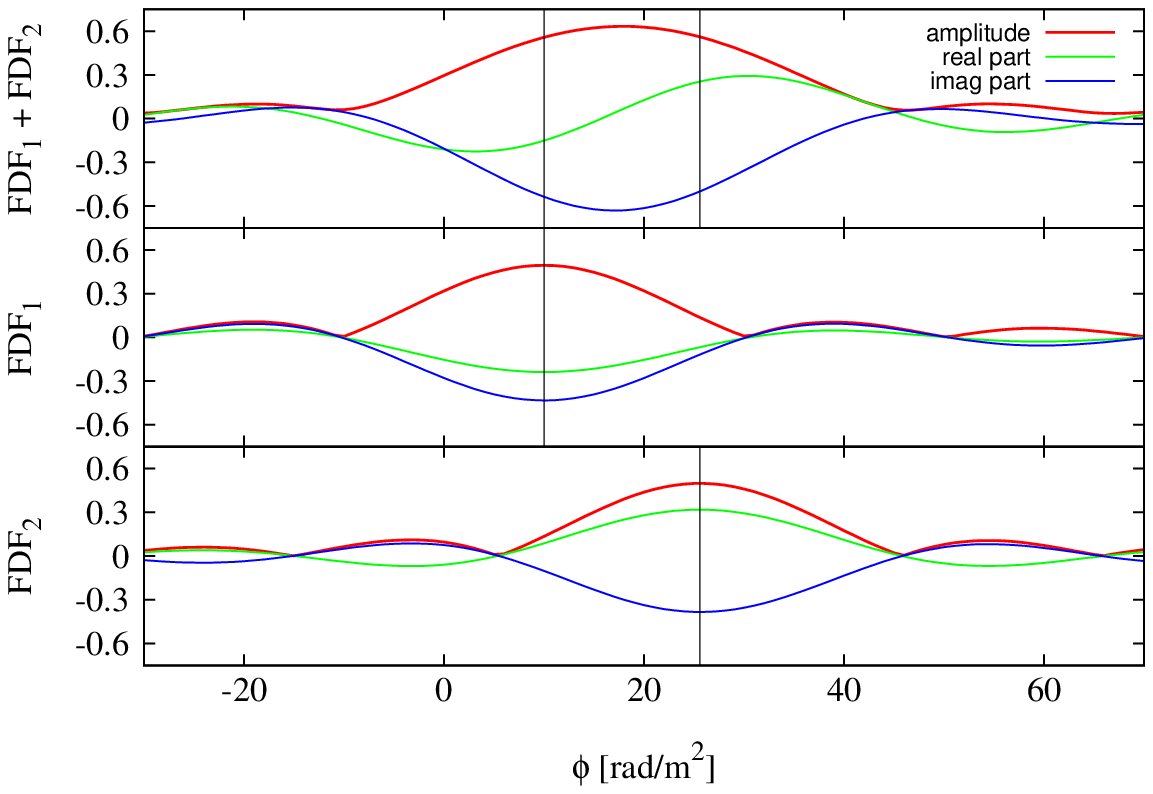} 
    \end{center}
   \end{minipage}
   \begin{minipage}{0.5\hsize} 
    \begin{center}
     \includegraphics[width=8.5cm]{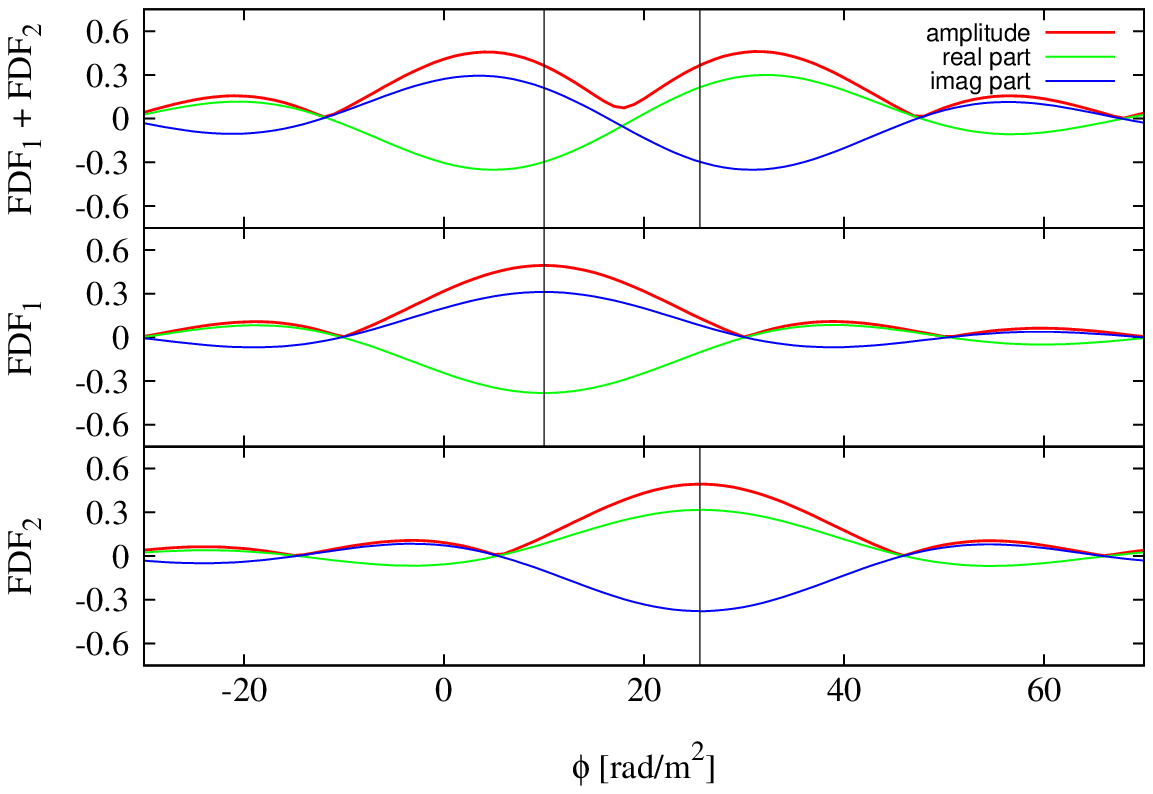} 
    \end{center}
   \end{minipage}
  \end{tabular} 
 \end{center}
\caption{The dirty Faraday Dispersion Functions (FDF) for $\Delta \phi = 0.7~{\rm FWHM}$ and $\Delta \chi_0 = 60^{\circ}$ (left) and for $\Delta \phi = 0.7~{\rm FWHM}$ and $\Delta \chi_0 = 10^{\circ}$ (right). The black lines show the correct location of the two sources. The red, green and blue line are the amplitude, real part and imaginary part of the dirty FDF, respectively. The middle and bottom panels represent the dirty FDFs in case that only the first ({FDF}$_1$) or second ({FDF}$_2$) source exists, and the top panel shows the sum of the dirty FDFs of the two sources ({FDF}$_1$+{FDF}$_2$).}
\label{fig:amb}
\end{figure*}

In order to understand the RM ambiguity more visually, we show Fig. \ref{fig:amb} which compares the dirty FDFs for ($\Delta \phi$,$\Delta \chi_0$) = (0.7 FWHM, 70 deg.) and (0.7 FWHM, 10 deg.), with which one and two sources are identified, respectively. One can see that the main peaks of the two dirty FDFs interfere with each other, and that RM ambiguity (does not) occurs when the signal from each source is constructive (destructive) between the two sources. Thus, the occurrence of the RM ambiguity depends on both the gap and initial polarization-angle difference. In addition, Fig. \ref{fig:amb} shows that RM ambiguity will be universal for any methods of RM synthesis, not just RM CLEAN.

\subsection{parameter estimation}

Next, we examine the estimation of the source parameters, that is, Faraday depth, amplitude, and intrinsic polarization angle, from the results of RM CLEAN. When two sources are identified, we estimate these parameters from the CLEAN components $S(\phi)$ as follows. First of all, in $\phi$ space, we regard CLEAN components in a beam centered on a peak of $S(\phi)$ with a width of the FWHM of the RMSF as contributing to the same source. This treatment comes from the fact that the resolution in $\phi$ space is roughly the FWHM of the RMSF and finer structure cannot be resolved. Then, the Faraday depth of a source is estimated as,
\begin{equation}
\phi_{\rm est} = \frac{\sum_k^{N_{\rm it}} |S^k_{\rm F}(\phi)| \phi} {\sum_k^{N_{\rm it}} |S^k_{\rm F}(\phi)|},
\end{equation}
where the sum is taken in the beam associated with the source in the above way. Secondly, the amplitude is estimated by the sum of the absolute values in each beam:
\begin{equation}
f_{\rm est} = \sum_k^{N_{\rm it}} |S^k_{\rm F}(\phi)| .
\end{equation}
Finally, we estimate the intrinsic polarization angle by,
\begin{equation}
\chi_{0, {\rm est}}
= \frac{1}{2} \tan^{-1} \left( \frac{{\rm Im}[\sum_k^{N_{\rm it}} S^k_{\rm F}(\phi)]}{{\rm Re}[\sum_k^{N_{\rm it}} S^k_{\rm F}(\phi)]} \right) - \phi_{\rm est} \lambda_0^2 ,
\end{equation}
where $\lambda_0^2$ is defined by Eq. (\ref{eq:lambda0}).

\begin{figure*}[htbp]
\begin{center}
\includegraphics[width=8cm]{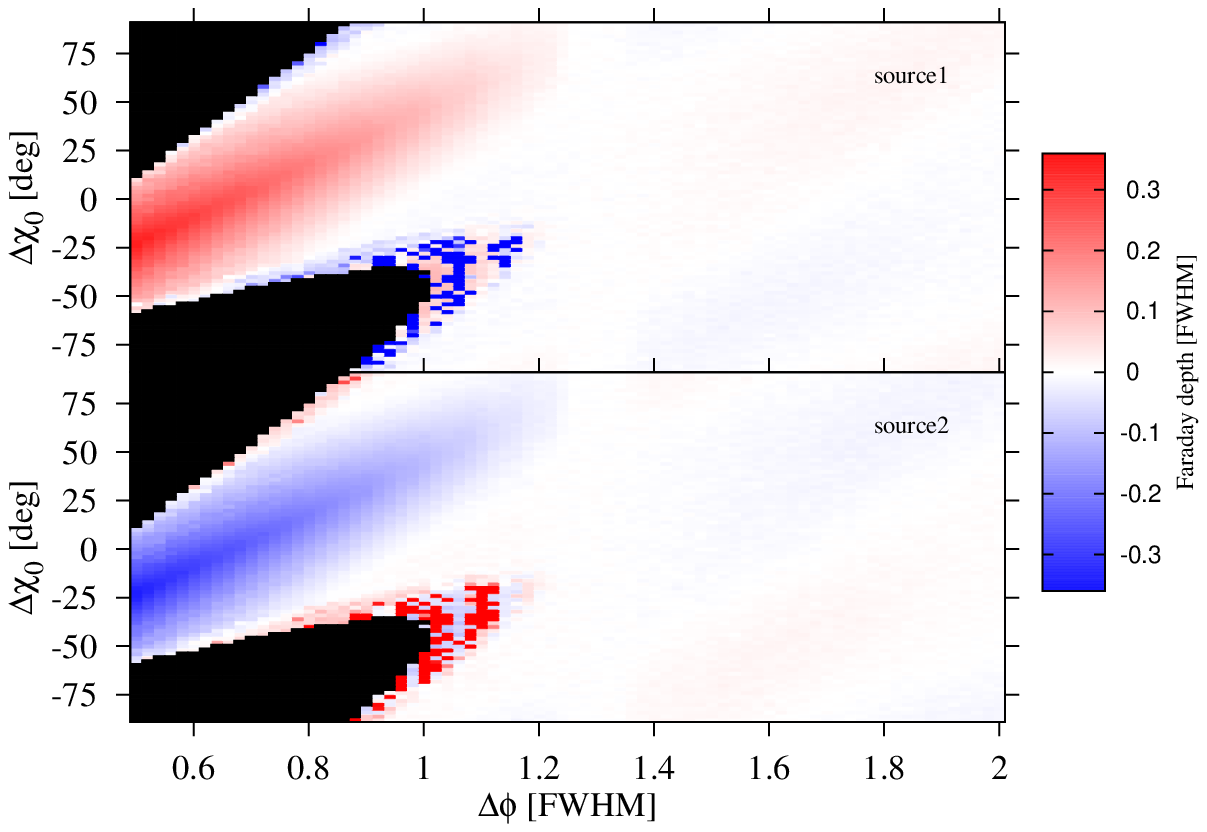} 
\includegraphics[width=8cm]{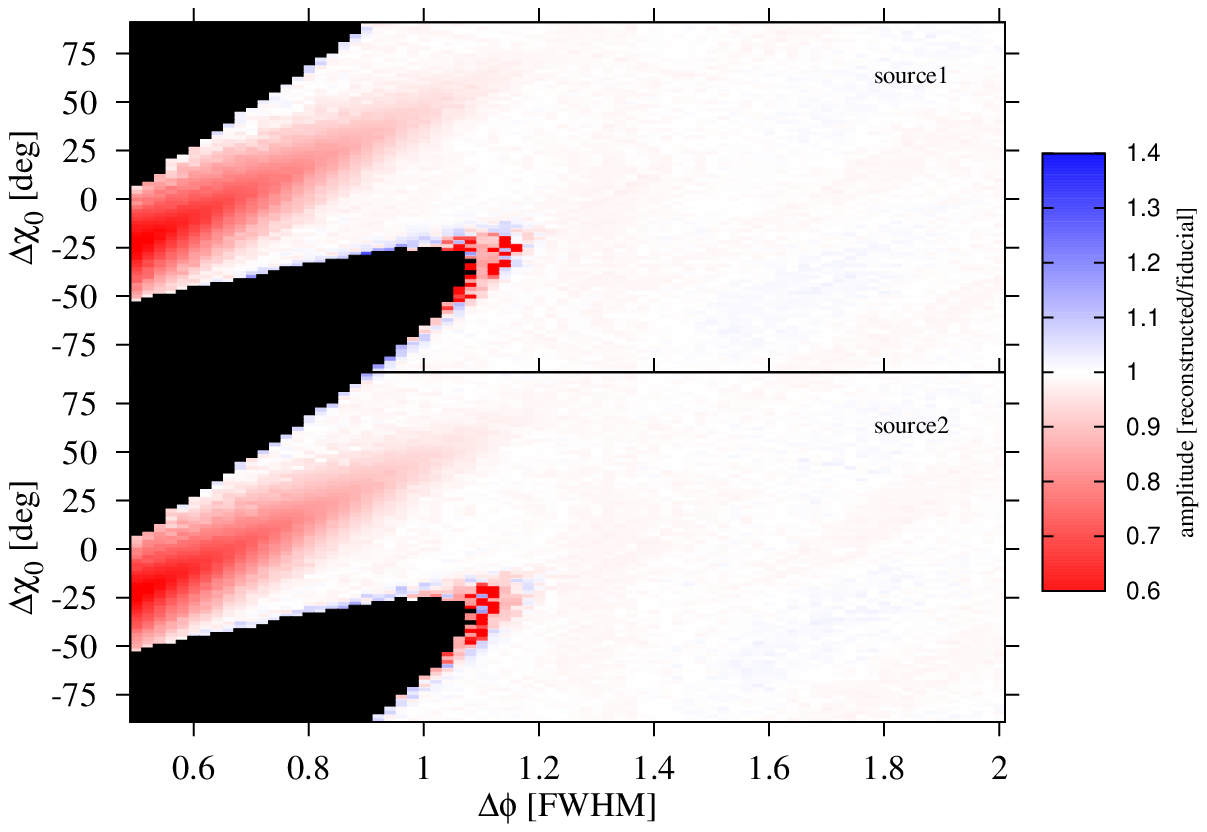}
\includegraphics[width=8cm]{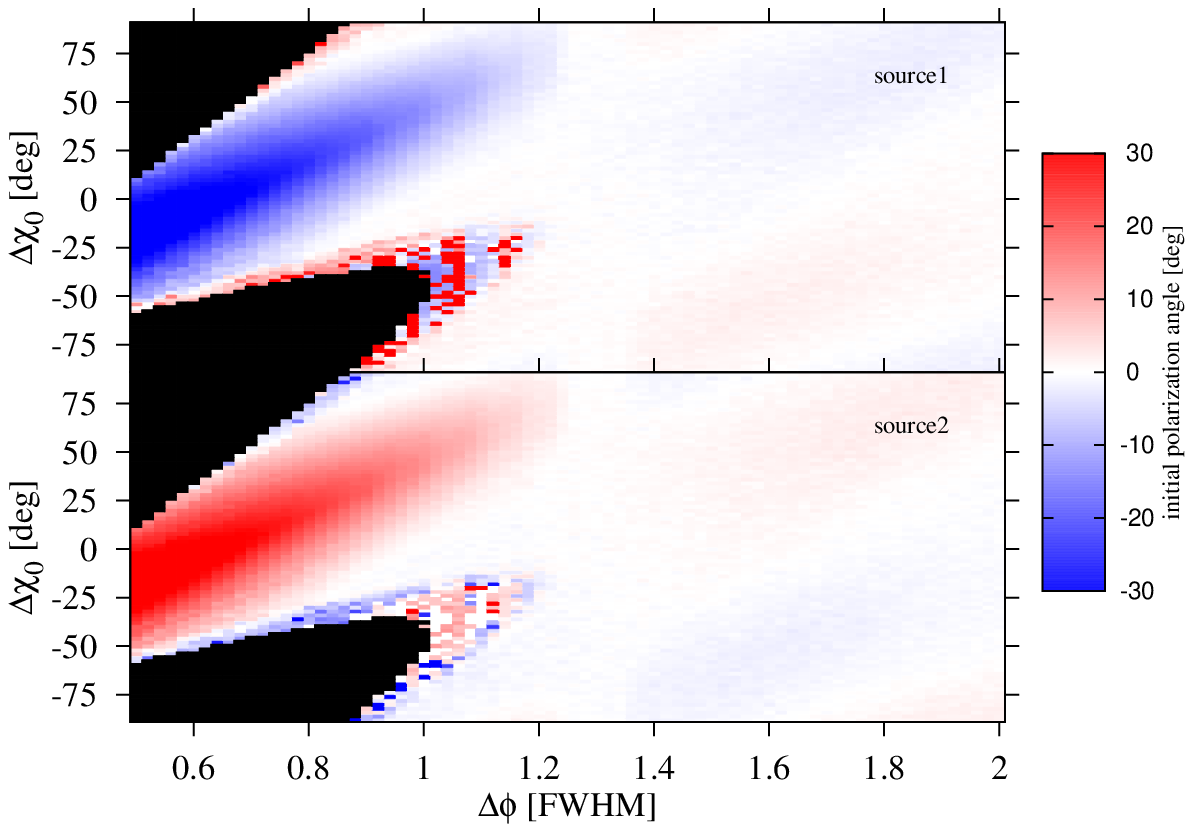}
\end{center}
\vspace{1cm}
\caption{Difference of the estimated and true values in $\Delta \phi$-$\Delta \chi_0$ plane for Faraday depth (top left), amplitude (top right) and initial polarization angle (bottom). Red (blue) region corresponds to underestimation (overestimation). The region where only one source is identified is masked by black. Two panels for each figure correspond to the two sources, respectively.}
\label{fig:para1}
\end{figure*}

Fig. \ref{fig:para1} shows the difference of the estimated and true parameter values in $\Delta \phi$-$\Delta \chi_0$ plane. Red (blue) region corresponds to underestimation (overestimation) of the parameter. The region where only one source is identified is masked by black. We calculate the estimation error for $\Delta \phi \geq$ 0.5 FWHM, because CLEAN component can be mixed with another source having a width of FWHM, and our method for parameter estimation cannot be utilized for $\Delta \phi \leq$ 0.5 FWHM. We see that the parameter estimation is relatively poor for $\Delta \phi \lesssim 1.2~{\rm FWHM}$. For $\Delta \phi \gtrsim 1.2~{\rm FWHM}$, the parameters are estimated very well and it is interesting to note that a kind of interference pattern is seen.

Let us see more details. The top left figure represents the errors in Faraday depths and we see the errors for the two sources are anti-correlated. In particular, for $\Delta \phi \lesssim 1.2~{\rm FWHM}$, the Faraday depth of one source with smaller $\phi$ $(= 10~{\rm rad/m^2})$ tends to be underestimated, while that of the other source tends to be overestimated. The errors can be as large as $\sim 8~{\rm rad/m^2} \sim 0.36~{\rm FWHM}$. This means that the gap of the two sources in $\phi$ space tends to be overestimated by as large as $\sim 0.72~{\rm FWHM}$ and we can confirm this from Fig. \ref{fig:amb}.

The top right figure of Fig. \ref{fig:para1} shows the errors in the amplitude of the two sources. They are mostly underestimated and the errors can be as large as $40\%$ of the true value. The bottom figure represents the initial polarization angles and they are anti-correlated.

Finally, we would like to emphasize that the parameter estimation errors shown here are mostly the intrinsic property of RM CLEAN and are not attributed to observation errors.

\subsection{chi-square analysis}

We have seen the performance of RM CLEAN in the previous subsections and found that it works well for $\Delta \phi \gtrsim 1.2~{\rm FWHM}$. Two sources can be resolved even for $\Delta \phi \lesssim 1.2~{\rm FWHM}$ depending on the relative initial polarization angle, although the parameter estimation is relatively poor. Nevertheless, because we cannot know the correct answer in the real observation, when we identify one source as a result of RM CLEAN, we cannot distinguish the two possibilities: (1) two sources are merged due to RM ambiguity or (2) there is truly only one source. Further, even if we can resolve two sources, we cannot know if the parameter estimation is reasonable or not.

Thus, we consider a possibility that the chi-square value of the fit between the observed polarization spectrum and that calculated from the result of RM CLEAN can be an indicator of the performance of RM CLEAN. The reduced chi-square is defined as,
\begin{equation}
\chi^2_\nu = \sum_{i = 1}^{N_{\rm ch}} \frac{1}{\sigma^2 \mu}
\left[ P_{\rm CLEAN}(\lambda^2_i) - P_{\rm obs}(\lambda^2_i) \right]^2,
\end{equation}
where $N_{\rm ch}$ is the number of channels, $P_{\rm obs}$ is the mock observation data of the polarization spectrum, $\sigma^2$ is the variance of the observation error, and $\mu$ is the number of data. In this calculation, $N_{\rm ch}$ is 1,100 and $\mu$ is 2,200 considering 2 Stokes parameters, Q and U. $P_{\rm CLEAN}$ is calculated from the CLEAN component using Eq. (\ref{eq:P}) as,
\begin{equation}
P_{\rm CLEAN}(\lambda^2) = \int_{-\infty}^{\infty} S(\phi) e^{2i \phi \lambda^2 } d\phi.
\end{equation}

\begin{figure}[t]
\begin{center}
\includegraphics[width=8cm]{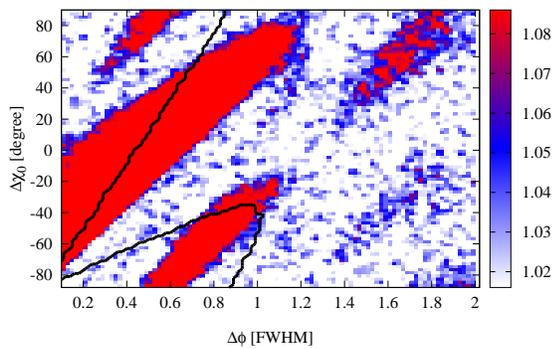} 
\end{center}
\caption{Reduced chi-square in $\Delta \phi$-$\Delta \chi_0$ plane. Parameter regions with a value of $\chi^2_{\nu}$ over 3-$\sigma$ (1.086) are colored in red. The thick black lines show boundary lines between the blacked and the white regions in Fig. \ref{fig:sourcenum}.
}
\label{fig:chisquare}
\end{figure}

Fig. \ref{fig:chisquare} shows the reduced chi-square, $\chi^2_{\nu}$, in $\Delta \phi$-$\Delta \chi_0$ plane. For the current number of datas, 3-$\sigma$ of $\chi$-square distribution corresponds to $\chi^2_{\nu} = 1.086$ and is colored in red. For $\Delta \phi \lesssim 1.2~{\rm FWHM}$, parameter regions with a value of $\chi^2_{\nu}$ over 3-$\sigma$ occupy a significant fraction of the plane and a similar pattern as in Figs. \ref{fig:sourcenum} and \ref{fig:para1} can be seen. But the pattern is slightly shifted in the $\Delta \chi_0$ direction and the red region does not exactly correspond to the black region in Fig. \ref{fig:sourcenum}. For $\Delta \phi \gtrsim 1.2~{\rm FWHM}$, most regions are less than 2-$\sigma$ level but red region can also be seen for posivive $\Delta \chi_0$. Finally, we have checked that the qualitative features do not change for larger values (20 and 30) of $f_1$ and $f_2$.


\subsection{Discussion} 

Here, we consider if the chi-square value calculated in the previous subsection can be an indicator of the performance of RM CLEAN. Comparing Figs. \ref{fig:sourcenum} and \ref{fig:chisquare}, we see that the regions where RM ambiguity occurs and where the reduced chi-square is over 3-$\sigma$ do not coincide with each other. This is because there are parameter sets $(\Delta \phi, \Delta \chi_0)$ with which one of the followings occurs:
\begin{enumerate}
\item[(i)] the fit of polarization spectrum is poor even though two sources are correctly resolved,
\item[(ii)] the fit of polarization spectrum is good even though two sources are not resolved.
\end{enumerate}
For the case (i), as can be seen from Fig. \ref{fig:para1}, the parameter estimation is relatively poor in the corresponding region. This would be the reason why the fit of the polarization spectrum is poor. 

\begin{figure}
\begin{center}
\includegraphics[width=7cm]{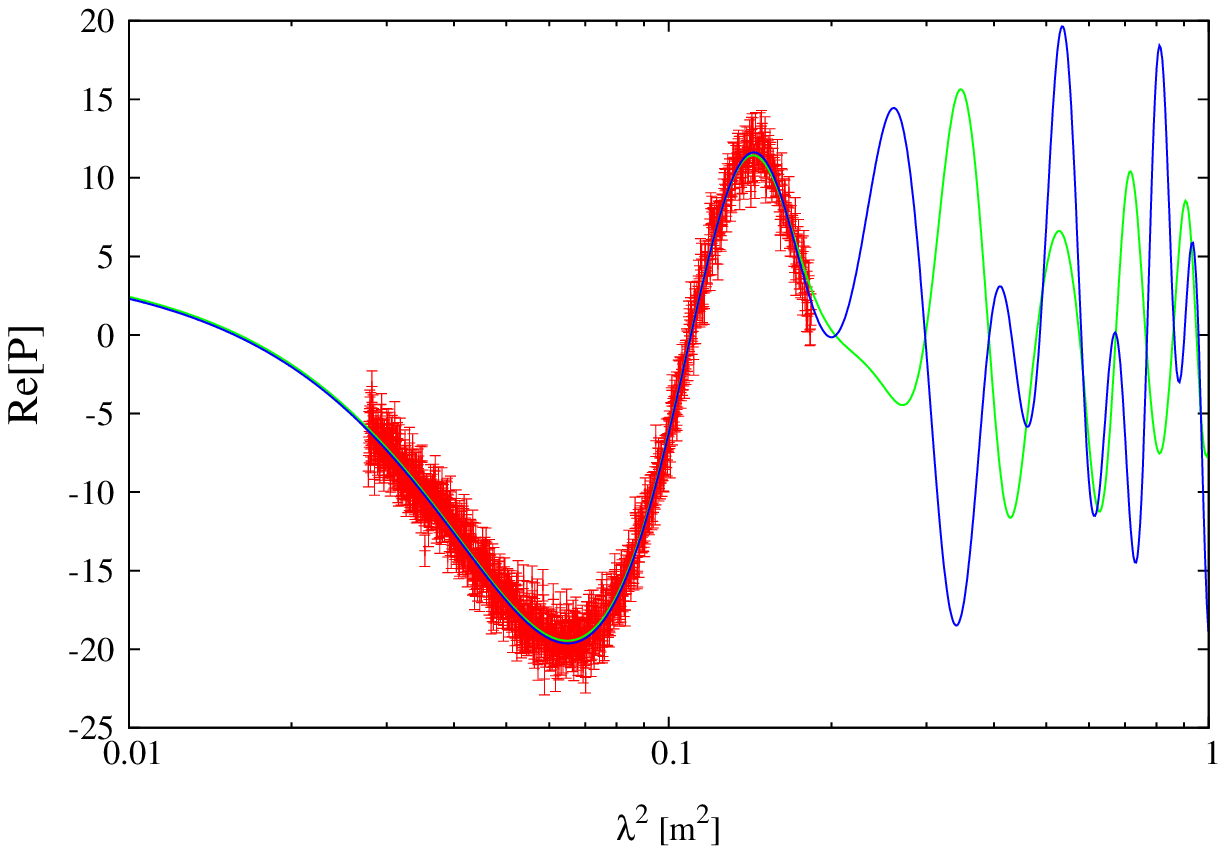} 
\includegraphics[width=7cm]{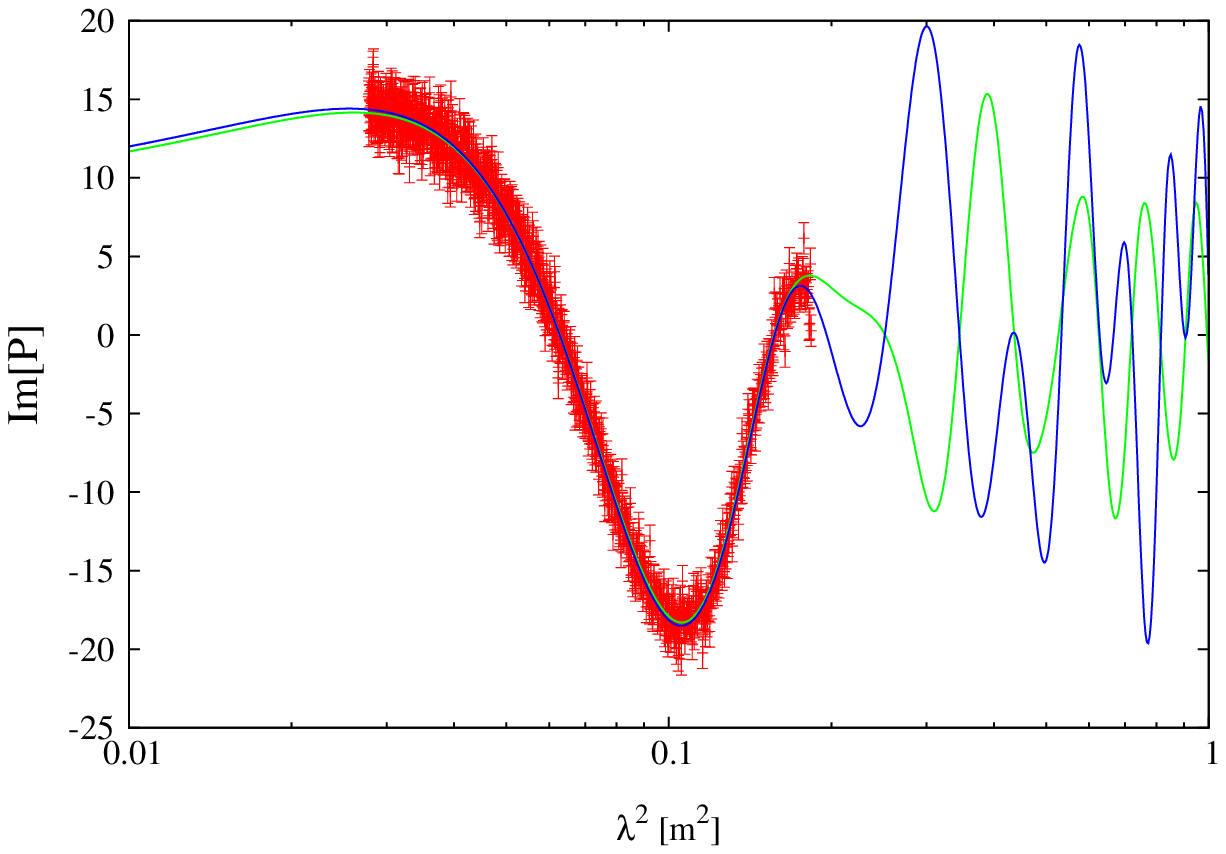} 
\end{center}
\caption{Polarization spectrum ($Q$ (top) and $U$ (bottom)) with $\Delta \phi = 0.6~{\rm FWHM}$ and $\Delta \chi_0 = 60^{\circ}$. The points with an error bar are mock data in the ASKAP band. The blue and green curves represent the polarization spectra calculated from the correct FDF and from the CLEAN components obtained by RM CLEAN, respectively. Only one source is identified for the values of $\Delta \phi$ and $\Delta \chi_0$ by RM CLEAN.}
\label{fig:logp}
\end{figure}

Next, let us consider the case (ii), which is more serious when we use the reduced chi-square to evaluate the performance of RM CLEAN. Fig. \ref{fig:logp} is a comparison of the polarization spectra calculated from the correct FDF and from the CLEAN components obtained by RM CLEAN for $\Delta \phi = 0.6~{\rm FWHM}$ and $\Delta \chi_0 = 60^{\circ}$. In this case  only one source is identified, although the reduced chi-square is very close to unity. We can see that the two polarization spectra coincide perfectly for the ASKAP band, while they deviate significantly from each other for longer wavelengths. Therefore, these two FDFs cannot be distinguished in the ASKAP band and shorter wavelengths even by ideal observations with no observational errors. It should be noted that this phenomenon is not a problem solely for RM CLEAN but is common for any algorithms. Thus, even if we identify only one source by RM CLEAN and the fit is good in the real observation, this does not always imply that the result is correct and there is a possibility that two (or more) sources located within $\sim 1.2~{\rm FWHM}$ are merged.

\section{Summary}

In this paper, we examined the performance of RM CLEAN by simulating spectro-polarimetric observations of two compact sources located in the same LOS. The observation noise was assumed to be relatively small to see the intrinsic properties of RM CLEAN. We varied systematically the relative initial polarization angle and Faraday depth. Especially, we focused on if the two polarized sources can be resolved in the Faraday depth space and how well the source parameters, such as the Faraday depth, emissivity and initial polarization angle, can be estimated. We confirmed the existence of RM ambiguity found in the previous studies. This is a phenomenon that two sources are merged and only one source is identified when they are closely located in the Faraday depth space ($\Delta \phi \lesssim 1.2~{\rm FWHM}$) for specific values of the relative initial polarization angle. The parameter estimation was also found to be poor for $\Delta \phi \lesssim 1.2~{\rm FWHM}$, even if two sources are identified. Further, we calculated the chi-square value for the fit between the mock data of polarization spectrum and the one from RM CLEAN. Then we found that the chi-squared value is not always significantly large even when RM CLEAN gives wrong results. This makes the standard chi-square analysis less reliable when using RM CLEAN.

\begin{ack}
K.T. is supported by Grand-in-Aid from the Ministry of Education, Culture, Sports, and Science and Technology (MEXT) of Japan, No. 24340048, No. 26610048 and No. 15H05896. S.I. was supported by the National Research Foundation of Korea through grant 2007-0093860.
\end{ack}


\end{document}